\documentclass[a4paper,12pt]{article}
\usepackage{jheppub_X} 
\usepackage{lineno}

\usepackage{amsmath}
\usepackage{amssymb}
\usepackage{setspace}
\usepackage{comment}

\def\a{\alpha}

\def\b{\beta}
\def\c{\chi}

\def\d{\delta}

\def\f{\frac}

\def\l{\left}

\def\la{\langle}
\def\ra{\rangle}

\def\mc{\mathcal}

\def\m{\mu}

\def\p{\partial}
\def\vp{\varphi}

\def\r{\right}

\def\vp{\varphi}

\def\be{\begin{equation}}
\def\ee{\end{equation}}

\def\bae{\begin{equation}\begin{aligned}}
\def\eae{\end{aligned}\end{equation}}
\def\bea{\begin{eqnarray}}
\def\eea{\end{eqnarray}}

\def\ba{\begin{array}}
\def\ea{\end{array}}

\definecolor{colorTC}{rgb}{.2,.7,.2}

\newcommand{\gap}{\hspace{0.8pt}}

\preprint{CERN-TH-2025-036}

\title{Superadditivity at Large Charge}

\author[a,b,c]{Timothy Cohen,}
\author[b,d ]{Ipak Fadakar,}
\author[b]{Andrew Gomes,}
\author[e]{\\[5pt] Alexander Monin,}
\author[b]{and Riccardo Rattazzi\,}

\affiliation[a]{\fontsize{10}{10}\selectfont Theoretical Physics Department, CERN, 1211 Geneva, Switzerland}
\affiliation[b]{\fontsize{10}{10}\selectfont Theoretical Particle Physics Laboratory, EPFL, 1015 Lausanne, Switzerland}
\affiliation[c]{\fontsize{10}{10}\selectfont Institute for Fundamental Science, University of Oregon, Eugene, Oregon 97403, USA}

\affiliation[d]{\fontsize{10}{10}\selectfont Scuola Normale Superiore, Piazza dei Cavalieri 7, Pisa, 56126, Italy}
\affiliation[e]{\fontsize{10}{10}\selectfont University of South Carolina, 712 Main St, 404, Columbia, South Carolina 29208, USA}

\emailAdd{tim.cohen@cern.ch, ipak.fadakar@sns.it, andrew.gomes@epfl.ch, amonin@mailbox.sc.edu, riccardo.rattazzi@epfl.ch}

\abstract{
The weak gravity conjecture has been invoked to conjecture that the dimensions of charged operators in a CFT should obey a superadditivity relation (sometimes referred to as convexity).  In this paper, we study superadditivity of the operator spectrum in theories expanded about the semi-classical saddle point that dominates correlators of large charge operators.  We explore this in two contexts.  The first is a model with two scalar fields that carry different charges, at a non-trivial Wilson-Fisher fixed point.  A careful analysis of the semi-classics for this two field model demonstrates that `quantum' violations of superadditivity (those not forbidden by the conjecture) persist in the large charge regime.  We then turn to study the general properties of CFTs at large charge as bottom-up EFTs. By a trial and error procedure we come up with a seemingly consistent family of examples violating the conjecture. In so doing the presence of a genuine dilaton field appears necessary. On the one hand our result demonstrates that the superadditivity conjecture cannot be proven purely on the basis of a  bottom-up analysis. On the other hand, the need for a dilaton, with the corresponding infinite fine tuning, indicates the conjecture-violating EFTs are unlikely to be UV completable. 
}

\begin{document}
\maketitle
\setcounter{page}{2}
\flushbottom

\newpage
\section{Introduction}
Much of what we know about Quantum Field Theory comes from studying perturbative amplitudes with a small number of legs.  The perturbation series defining such amplitudes are well-known to be asymptotic: at some perturbative order the expansion breaks down due to a factorial growth in the number of Feynman diagrams~\cite{rubakov1995nonperturbative}. This order, known as the point of optimal truncation, is generically $\mathcal{O}(1/\lambda)$ where $\lambda$ is a proxy for the perturbative couplings in the model. This means that for weak coupling, perturbation theory is usually sufficiently precise for making predictions. However, sometimes the problem appears to be much worse. When considering processes with a large number $n$ of external legs, the optimal truncation appears to arrive earlier in the series~\cite{Zinn-Justin:2002ecy}. Indeed once $\lambda n \gtrsim 1$ even the tree level amplitude cannot be trusted.
 
Such a paradoxical situation where perturbation theory breaks down despite weak coupling originates from having expanded the theory around the wrong semi-classical saddle in the path integral, namely the vacuum field configuration.  Indeed, it should be possible to reorganize calculations for amplitudes with a large number of legs by expanding about a non-trivial saddle point. A first proper treatment of this saddle point was performed in the context of $\lambda \phi^4$ theory~\cite{Son_1996}. Intuitively, the large number of field insertions can no longer be thought of as small perturbations around the trivial saddle, and should instead be `pulled into' the exponent of the path integral kernel $e^{-S}$, which can be interpreted as a modified effective action. This strategy has led to a plethora of interesting results, for example the recent numerical result~\cite{Demidov_2023}.

For theories with a conserved charge, considering operators with large global charge (for example, the operator $\phi^n$ with $n \gg 1$ in the $\lambda |\phi|^4$ theory) can provide a useful theoretical handle. Analytic progress has been made in analyzing the large charge expansion of field theories, see \emph{e.g.}~\cite{Hellerman_2015,Hellerman_2017,Monin_2017,Badel_2019,Badel_2023,Jafferis_2018,Antipin_2022,Antipin:2021rsh}. When the theory of interest is a Conformal Field Theory (CFT) with a $U(1)$ global symmetry, calculations can be performed  systematically by relying on the state-operator correspondence.  In particular, the CFT large charge operator spectrum and correlators can be computed as a systematic expansion in inverse powers of the corresponding charge.

Our goal in this paper is to apply the techniques and general lessons learned from studying CFTs at large charge to explore the `Abelian Convex Charge Conjecture'~\cite{Aharony_2021}:
\be\label{conv}
\Delta(n_1 q_0 + n_2\gap q_0) \geq \Delta(n_1 q_0) + \Delta(n_2\gap q_0) \, ,
\ee
where $\Delta(q)$ is the minimal dimension of an operator of charge $q$ in the CFT, $q_0$ is declared to be some $\mathcal{O}(1)$ multiple of the minimal charge, and the $n_i$ are positive integers.  Since this is a statement about discrete quantities, we will refer to this as the `Superadditivity Conjecture.'  The motivation for \eqref{conv} comes from the Weak Gravity Conjecture (WGC), as we will review below.  In particular, this conjecture is also expected to hold when the integers $n_1$ and $n_2$ are large, and so clearly the large charge expansion is relevant to exploring the validity of \eqref{conv}.

We will first clarify the sense in which the conjecture holds by studying a detailed example of a two field CFT in the $\epsilon$-expansion.  This will demonstrate the role of the choice for $q_0$ in \eqref{conv} to ensure that the superadditivity conjecture holds.  We will then turn to a bottom up analysis, where we will give a concrete example of a low energy EFT that could in principle lead to superadditivity violation.  We do not know how to UV complete this EFT, so it could well only exist in the swampland.  Nonetheless, it provides a concrete target to eliminate on the quest to find a proof of the superadditivity conjecture.

The rest of this paper is organized as follows.  In Section~\ref{review}, we review the conjecture \eqref{conv} and its motivation by the WGC, as well as progress in understanding the spectra of CFTs with global charge.  In Section~\ref{micro}, we consider a model in which $q_0$ is not the minimal charge of the theory, explaining how this `microscopic non-convexity' manifests at large charge and what features appear to be universal. In Section~\ref{eft}, we review the large charge superfluid EFT and then use it to construct conjecture-violating EFTs, commenting on the implications. Section~\ref{disc} concludes this paper with a discussion of the lessons to be drawn from these examples.

\section{Superadditivity From Weak Gravity}\label{review}
In this section, we briefly review the argument of Ref.~\cite{Aharony_2021} for the `Abelian Convex Charge Conjecture' (what we will call the `Superadditivity Conjecture') given in \eqref{conv}.  The Weak Gravity Conjecture (WGC) \cite{Arkani_Hamed_2007} states that for any UV complete theory of quantum gravity with a $U(1)$ gauge symmetry, there must exist a particle whose charge is greater than its mass in Planck units.  In other words, gravity must be the weakest force in nature.  If one were to place two particles with this minimal charge beside each other, their electric repulsion would overcome their gravitational attraction, thereby preventing the formation of a bound state.  This leads to a natural generalization of the WGC, the `Positive Binding Conjecture' (PBC) \cite{Palti_2017,Heidenreich_2019}, which states that there should exist a particle whose self-binding energy is non-negative.

Importantly, the PBC differs from the WGC when there are long-range forces mediated by light scalar fields \cite{Aharony_2021}. For example, Ref.~\cite{Fitzpatrick:2011hh} considered the theory of a complex scalar $\phi$ in $\text{AdS}_5$, charged under a $U(1)$ gauge symmetry, and with quartic interactions. The self-binding energy
\be
\gamma_{\phi^2} \equiv \Delta_{\phi^2} - 2 \Delta_\phi \, ,
\ee
where $\Delta$ denotes the scaling dimension of the corresponding operator in the dual $\text{CFT}_4$, is a sum of three terms:
\be
\gamma_{\phi^2} =  \gamma_{\phi^2}^\text{photon}+\gamma_{\phi^2}^\text{gravity}+\gamma_{\phi^2}^\text{quartic} \, .
\ee
Only the sum of these terms is positive. Thus, Ref.~\cite{Aharony_2021} proposed a natural CFT dual statement to the PBC: the so-called Abelian Convex Charge Conjecture \eqref{conv}.
As stated before, since it is a condition on discrete quantities, it should more precisely be referred to as superadditivity. While for functions $\Delta \propto q^\alpha$ convexity and superadditivity are the same, this is not always the case. For instance, the function $q-\frac{1}{q}$ is superadditive but concave, while $q+\frac{1}{q}$ is convex but subadditive.

Two clarifications of the conjecture are in order. First, $q_0$ need not be the minimal charge of the theory. The reason, as pointed out in Ref.~\cite{Aharony_2021}, is that such a charge may not have the minimal dimension-to-charge ratio. For example, consider the simplest interacting Wess-Zumino model in 4d:
\be
W = g\gap \Phi^3 \, ,
\ee
whose scalar and fermion fields have $U(1)_R$ charges of $2/3$ and $-1/3$, respectively. The IR fixed point of this theory is a free theory ($g \rightarrow 0$), implying that the dimensions of these fields at the fixed point are $1$ and $3/2$, respectively. Trivially then, $\Delta(2/3) < 2 \, \Delta(1/3)$. One can easily show that the conjecture is instead valid in this case for $q_0 = 2/3$.

The second clarification concerns the need for $q_0$ to be order one. In the original conjecture it was an acknowledged assumption that an interacting CFT (excluding those CFTs with moduli spaces)\footnote{\label{footFlat}Other than in free theories, the only known way for the conjecture to be exactly saturated is with supersymmetry. Suppose that there exists a chiral operator $\mathcal{O}$ in the spectrum. In this case,
\be
\Delta(\mathcal{O}^n) = n \gap \Delta(\mathcal{O}) \, .
\ee
It is also possible to construct non-supersymmetric CFTs with flat directions in the large $N$ limit~\cite{PhysRevD.36.562,PhysRevLett.52.1188,Chai:2020onq,Chai:2020zgq}.} would have $\Delta(q) \sim q^{d/(d-1)}$ for $q$ large, which is manifestly convex (see Section~\ref{eft} for an explanation of this scaling). Therefore, it was believed that the conjecture is trivially satisfied except in the case that $q_0$ was order one. However, a family of counterexamples was demonstrated in Ref.~\cite{Sharon_2023}, where $q_0$ could be made an arbitrarily large multiple of the minimal charge. These authors constructed a supersymmetric, clockwork-like Chern-Simons model with fields of charge $1, 3, \dots, 3^N$ for arbitrary $N$. This model violates superadditivity for all $q_0 < \mathcal{O}(3^N)$, the lesson being that large $N$ can delay the onset of convexity.\footnote{It was speculated that it may be possible to place a bound on $q_0$ given some measure of the number of degrees of freedom, however no concrete bound was proposed.}

Therefore, it now seems that the content of the conjecture is not in $q_0$ being order one, but rather in the assumption of universal $\Delta(q) \sim q^{d/(d-1)}$ behavior for (non-supersymmetric) interacting CFTs. This seems plausible and there are no known UV complete counterexamples, but no proof is known.  We will give large charge EFT counterexamples below in Section~\ref{eft}, although we do not know how to UV complete them and so do not interpret them as providing counterexamples.

\section{Superadditivity and Quantum Fluctuations}\label{micro}
In this section, we explore the role of the $q_0$ factor in \eqref{conv} in the context of a simple Wilson-Fisher CFT with two scalar fields that carry different charges under a global $U(1)$.  This model has a non-trivial interacting fixed point in the $\epsilon$-expansion.  We can thus apply the semi-classical techniques developed in Ref.~\cite{Badel_2019} to this example, which will allow us to check superadditivity explicitly.  We will confirm that superadditivity does not hold for general $q_0$ due to the presence of quantum excitations.  However, there does exist a choice of $q_0$ such that \eqref{conv} holds.  This example also provides an opportunity to illustrate for the first time the semi-classics of the large charge expansion in a theory with multiple scalars of different charges.  In particular, we will show that no new stable saddle points exist, beyond the one that determines the physics of the single field theory at large charge.

\subsection{A Two-field Model}
\label{sec:TwoFieldModel}
The model contains two complex scalar fields $\phi_1$ and $\phi_2$.  The Lagrangian is 
\begin{equation}
\label{eq:lag}
    \mathcal{L} = \sum_{i = 1,2} \bigg( |\partial \phi_i |^2 + \frac{\lambda_i}{4} |\phi_i |^4 \bigg) + \kappa |\phi_1 |^2 |\phi_2 |^2 + \bigg(\frac{\eta}{6} \phi_1 \phi_2^{\dag 3} + \text{h.c.}\bigg)\,.
\end{equation}
We normalize the charges to $Q(\phi_1) = 1$ and $Q(\phi_2) = 1/3$, and we work in Euclidean signature where $d = 4-\epsilon$ dimensions throughout this section.  This theory has a Wilson-Fisher fixed point; for a general analysis of multi-scalar field models with Wilson-Fisher fixed points, see~\cite{Osborn:2020cnf}.

The properties of small charge operators in this theory already make the need to appropriately choose $q_0$ clear.  Recall that $\Delta(q)$ is the minimal dimension of an operator with charge $q$.  At tree-level, we have 
\begin{align}
\Delta(1/3) = [\phi_2] = 1\,, \qquad
\Delta(2/3) = [(\phi_2)^2] = 2\,, \qquad
\Delta(1) = [\phi_1] = 1\,.  
\end{align}
If we choose $q_0 = 1/3$, then this clearly violates the superadditivity conjecture \eqref{conv}.  However, if we take $q_0 = 1$ then we never encounter the contribution from $\phi_2$ (since operators involving $\phi_2$ necessarily have non-minimal dimension), and the conjecture is satisfied.

We now show that the same story holds for correlators of large charge $q$ with
\be
\epsilon \gap q \gg 1
\ee
for the theory evaluated at its non-trivial Wilson-Fisher fixed point.  In particular, we will argue that the only non-trivial saddle that must be considered in the large charge limit is a function of $\phi_1$ alone.  The one-loop fixed point equations are 
\bae\label{fp}
    \bar{\lambda}_1 &= 5 \bar{\lambda}_1^2 + 4 \bar{\kappa}^2 \\[3pt]
    \bar{\lambda}_2 &= 5 \bar{\lambda}_2^2 + 4 \bar{\kappa}^2 + 2 \bar{\eta}^2 \\[3pt]
    \bar{\kappa} &= 4 \bar{\kappa}^2 + 2 \bar{\kappa} (\bar{\lambda}_1 + \bar{\lambda}_2) + \bar{\eta}^2 \\[3pt]
    \bar{\eta} &= 3 (2 \bar{\kappa} + \bar{\lambda}_2)\gap \bar{\eta} \, .
\eae
where we have defined
\be\label{barred}
\lambda_i = (16 \pi^2 \epsilon) \gap \bar{\lambda}_i \, ,
\ee
and likewise for $\kappa$ and $\eta$. The equations have a unique non-trivial solution\footnote{For general parameters the potential is unbounded from below.  However, for this solution it is non-negative.}
\bae
    \bar{\lambda}_1 \simeq 0.16\,, \qquad
    \bar{\lambda}_2 \simeq 0.15\,, \qquad
    \bar{\kappa} \simeq 0.093\,, \qquad
    \bar{\eta} \simeq 0.046 \, .
    \label{eq:fixedPointVals}
\eae
Notice the very approximate exchange symmetry of this solution: $\lambda_1 \simeq \lambda_2$ and $\eta < \lambda_i,\kappa$. Combined with the relatively small coefficients of $\bar{\eta}^2$ in \eqref{fp}, the exchange symmetric limit $\eta \to 0$ only leads to $\mathcal{O}(10^{-2})$ corrections.

To calculate the dimensions of charged operators at the fixed point, we follow the same semi-classical method as in Ref.~\cite{Badel_2019}. For our purposes it is sufficient to work at the leading order in the $\epsilon$-expansion. At the fixed point, we will exploit the operator-state correspondence to calculate scaling dimensions. To this end, we can map the theory to the Euclidean cylinder $\mathbb{R} \times S^{d - 1}$ with radius $R$, where the directions along the sphere are parameterized by $d-1$ coordinates $\vec{\theta}$. This has the effect of introducing a conformal mass $m = \frac{d-2}{2 R} \xrightarrow{\,d\gap =\gap 4\,} \frac{1}{R}$, so the Lagrangian in \eqref{eq:lag} picks up a mass term:
\begin{equation}\label{lagm}
    \mathcal{L} = \sum_{i = 1,2} \left( |\partial \phi_i |^2 + m^2 |\phi_i |^2 + \frac{\lambda_i}{4} |\phi_i |^4 \right) + \kappa |\phi_1 |^2 |\phi_2 |^2 + \bigg(\frac{\eta}{6} \phi_1 \phi_2^{\dag 3} + \text{h.c.}\bigg)\,.
\end{equation}

In order to find the semi-classical saddle of the path integral of large charge correlation functions, it is useful to perform the change of variables:
\begin{equation}\label{change}
    \phi_i = \frac{\rho_i}{\sqrt{2}} e^{i \chi_i} \, ,
\end{equation}
where the $\c_i$ fields are defined modulo $2\pi$, and this change of variables is only non-singular if $\rho_i \neq 0$. The Lagrangian in terms of these fields is given by
\begin{align}
    \mathcal{L} &= \sum_{i = 1,2} \bigg( \frac{1}{2} (\partial \rho_i )^2 + \frac{1}{2} \rho_i^2 (\partial \chi_i )^2 + \frac{1}{2} m^2 \rho_i^2 + \frac{\lambda_i}{16} \rho_i^4 \bigg) \notag\\[2pt]
    &\hspace{22pt} + \frac{\kappa}{4} \rho_1^2 \rho_2^2 + \frac{\eta}{12} \rho_1 \rho_2^3 \cos(\chi_1 - 3 \chi_2) \, .
\label{eq:CylinderPolarlagrangian}
\end{align}
The corresponding charge is
\be
\label{eq:U(1)Charge}
Q = \int \text{d}^{d-1} \theta \l ( \rho_1^2\gap \p_0\c_1 + \f{1}{3}\rho_2^2\gap \p_0\c_2 \r )  \, .
\ee

\subsection{The Semi-classical Analysis}

We now turn to finding the semi-classical solution that dominates the path integral for the correlators of large charge operators.  Since we are interested in the dimensions of large charge operators at the Wilson-Fisher fixed point, we can use the CFT operator-state correspondence to compute the minimal scaling dimension of a state with charge $q \in \mathbb{Z}/3$.  Specifically, we want to evaluate the correlator 
\be\label{evo}
\la \psi_q | e^{-HT}  | \psi_q  \ra \underset{T\to \infty}{\sim} e^{-E_q T}  \, ,
\ee
where $| \psi_q  \ra$ is a state of charge $q$, and $T$ is Euclidean time. If $| \psi_q  \ra$ is sufficiently generic (in particular if it overlaps with the charge $q$ state of minimal energy), the operator-state correspondence implies that 
\be
\Delta(q) = E_q R\,.
\label{eq:DeltaEq}
\ee
We may write a charge $q$ as $q = n_1 + n_2 / 3$, for $n_i \in \mathbb{Z}$. There are in general many choices for the $n_i$; we will discuss the interpretation of these choices below.  It is then convenient to consider the linear combinations 
\begin{subequations}\label{comb}
\begin{align}
    q\gap \chi &= n_1 \chi_1 + n_2 \chi_2 \\[3pt]
    \omega &= \chi_1 - 3 \chi_2 \, .
\end{align}
\end{subequations}
Note that while $\chi$ and the $U(1)$ invariant $\omega$ are convenient variables in what follows, the well-defined ($2\pi$ periodic) degrees of freedom are the $\chi_i$. The $U(1)$ charge operator \eqref{eq:U(1)Charge} can now be written as
\begin{align}
\label{eq:Qfunctional}
    Q &= -i \int_{S^{d-1}} \text{d}^{d-1}\theta \, \f{\d}{\d{\chi(\vec \theta)}} \, .
\end{align}
Thus, following Ref.~\cite{Badel_2019}, the charge $q$ state that we shall evolve in Euclidean time is
\begin{equation}\label{bd}
    | \psi_q  \ra = \int \mc D \chi \, \exp \l [ i \frac{q}{\Omega_{d-1} R^{d-1}
    } \int_{S^{d-1}} \text{d}^{d-1}\theta \, \chi(\vec{\theta}) \r ] |\rho_1=f_1, \rho_2=f_2, \chi(\vec{\theta}), \omega=\omega_0 \ra \, ,
\end{equation}
where the $f_i$ and $\omega_0$ are constants to be conveniently fixed later, and $\Omega_{d-1}$ is the total solid angle of the $S^{d-1}$ sphere.

Using \eqref{bd}, we can express the correlator in \eqref{evo} as a Euclidean path integral:
\be
\la \psi_q | e^{-HT}  | \psi_q  \ra = \mathcal{Z}^{-1} \int \mc D \phi_i\gap \mc D \Bar{\phi}_i\gap e^{-\int_0^T \text{d}\tau\gap \text{d}^{d-1}\theta\gap \mc L_\text{eff}} \, ,
\ee
where $\mathcal{Z}$ is the vacuum-to-vacuum correlator ensuring that $E_0 = 0$, and with
\be\label{eff}
\mc L_\text{eff} = \mc L + i \frac{q}{\Omega_{d-1} R^{d-1} } \dot{\chi} \, .
\ee
Due to the ansatz \eqref{bd}, this path integral is supplemented by the Dirichlet boundary conditions $\rho_i=f_i$ and $\omega=\omega_0$. The second term in \eqref{eff}, being a total derivative, leads to Neumann boundary conditions for $\chi$ that we account for below.  The leading approximation to this path integral comes from extremizing the effective action, leading to a non-trivial semi-classical saddle. We will tune the coefficients $f_i$ and $\omega_0$ so that this solution is stationary (constant Lagrangian) and therefore the action increases linearly in Euclidean time.  From this point of view, this solution to the equations of motion gives the tree-level contribution to the correlator, while from the point of view of the perturbation theory around the trivial vacuum, this is an all order resummation of the leading large $q$ loop corrections \cite{Badel_2019}.

We are interested in spherically homogeneous and stationary saddle-point solutions. The bulk equations of motion for the angular fields are
\begin{subequations}
\label{ang}
\begin{align}
\p_\tau (\rho_1^2 \dot{\chi}_1) &= -\eta \rho_1 \rho_2^3 \sin (\chi_1 - 3 \chi_2) \\[3pt]
\p_\tau (\rho_2^2 \dot{\chi}_2) &= 3 \eta \rho_1 \rho_2^3 \sin (\chi_1 - 3 \chi_2) \, .
\end{align}
\end{subequations}
In order for the solution to be stationary we must take
\be\label{anz}
\rho_i = f_i  \qquad\text{and}\qquad \omega = \omega_0 = k\gap \pi \, ,
\ee
which implies that the right hand side of \eqref{ang} is zero. The solutions for the $\chi_i$ are thus
\begin{align}
\label{eq: Saddle}
\chi_1= -i \m \, (\tau + c)  \qquad\text{and}\qquad \chi_2= - \frac{1}{3} i \m \, (\tau + c)\,,
\end{align}
where $\tau$ is Euclidean time, and the unconstrained constant $c$ reflects the superfluid-like symmetry breaking pattern $\mathbb{R}_\tau \times U(1) \rightarrow \mathbb{R}_\text{diag}$. The constant $\mu$ is fixed by employing the Neumann condition (boundary equation of motion from \eqref{eff}):
\begin{align}
\label{eq: Saddle2}
\mu \left( f_1^2 + \frac{1}{9}f_2^2 \right)   = \frac{q}{\Omega_{d-1} R^{d-1}} \, .
\end{align}
The constant $\mu$ plays the role of the chemical potential. Since there remains an effective time translation symmetry $\mathbb{R}_\text{diag}$, there is the possibility of writing down an effective Hamiltonian for the theory around the saddle point. As we show in Section \ref{sec:pert}, the effective theory is stable.\footnote{Furthermore, in Section \ref{sec:interp} we discuss the interpretation of the saddle points as Bohr-Sommerfeld orbits. Solutions with non-constant $\rho_i$ and $\omega$ are necessarily oscillating, and therefore correspond to states around this `ground state' with excited angular and radial modes.}

Plugging \eqref{eq: Saddle} into the equations of motion for the radial fields that come from the Lagrangian \eqref{eq:CylinderPolarlagrangian}, we have
\begin{subequations}
\label{eq:f1f2equations}
\begin{align}
     \left(\mu^2 - m^2 \right) f_1 &= \frac{\kappa}{2} f_2^2 f_1 + \frac{\lambda_1}{4} f_1^3 + (-1)^k \frac{\eta}{12} f_2^3 \label{eq:f1equation}\\[3pt]
     \left(\f{\mu^2}{9} - m^2 \right) f_2 &= \frac{\kappa}{2} f_1^2 f_2 + \frac{\lambda_2}{4} f_2^3 + (-1)^k \frac{\eta}{4} f_1 f_2^2\,,
\end{align}
\end{subequations}
where the $k$ dependence comes from \eqref{anz}.
For $m < \m < 3m$, there is only one real solution for which $f_2=0$; this corresponds to the one-field solution found in \cite{Badel_2019}. For $3m < \m$ and $k$ even, there exists another solution with both $f_i\neq 0$.  Note that since the $\cos(\chi_1-3\chi_2) = \cos\omega = 1$ when $k$ is even, this solution is a local maximum along the $\omega$ direction. By the approximate exchange symmetry this second (two-field) solution has a relatively small $f_1$; specifically, it is smaller than $f_2$ by a factor of $\mathcal{O}(10^2)$ in the limit $q \rightarrow \infty$.

\subsection{Interpretation of Solutions}\label{sec:interp}

Using \eqref{evo} and \eqref{eq:DeltaEq}, one can find the scaling dimensions associated to each of these solutions. For the one-field solution, an analytic expression was given in Ref.~\cite{Badel_2019}. Here we simply give the asymptotic behavior:
\be\label{sol1}
\Delta^{(1)} (q) \simeq \begin{cases} 
          q & \epsilon\gap q \ll 1 \\[3pt]
          \frac{6\pi^2}{\lambda_1} \left(\frac{\lambda_1 q}{8\pi^2} \right)^{4/3} & \epsilon\gap q \gg 1
       \end{cases} \, .
\ee
Note that since this solution has $f_2 = 0$, the change of variables (\ref{change}) for $\phi_2$ was singular. Maintaining the consistency of this change of variables for $\phi_2$ means that we must take $n_2 = 0$ in \eqref{comb} and thus $q = n \in \mathbb{Z}$.\footnote{Due to the gauge symmetry $\chi_1 \rightarrow \chi_1 + 2\pi$, taking $q \notin \mathbb{Z}$ in (\ref{bd}) results in the wavefunction becoming multi-valued.}

For later convenience, we also provide the large charge ($n \to \infty$) solutions for $\mu$ and $f_1$:
\begin{align}
\label{eq:muf}
\mu^{(1)} R = \left(\frac{\lambda_1 n}{8\pi^2} \right)^{1/3} \qquad\text{and} \qquad\,\, f_1^{(1)} R = \left(\frac{ n}{\pi^2 \lambda_1^{1/2}} \right)^{1/3} \, .
\end{align}
It is interesting to note that the gap between neighboring scaling dimensions, is given by the chemical potential:
\be\label{gap}
\Delta^{(1)}(n+1) - \Delta^{(1)}(n) \simeq \mu^{(1)} R \, .
\ee

\begin{figure}[htbp]
\centering
\includegraphics[width=\textwidth]{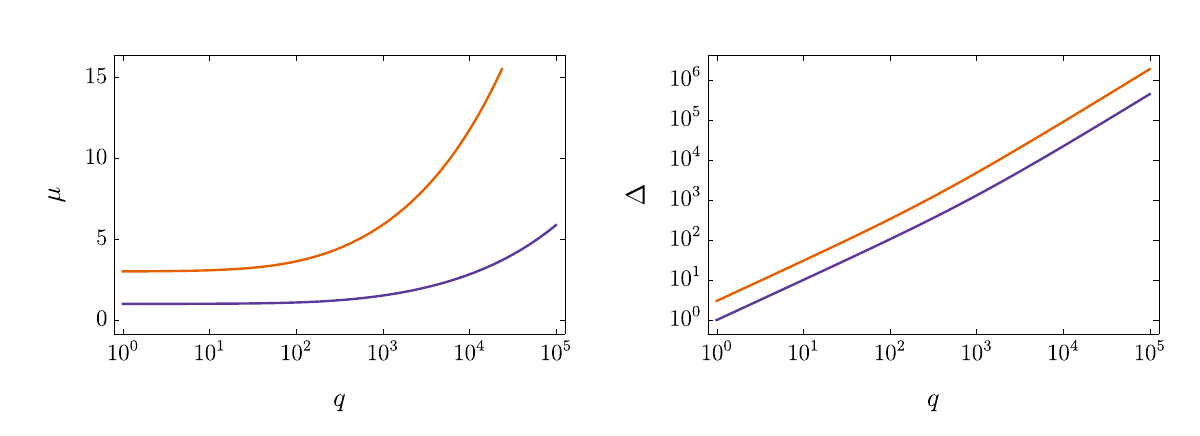}
\caption{ The chemical potential $\mu$ (left) and scaling dimension $\Delta$ (right) as functions of the charge $q$ for the two saddle points. In both cases the bottom (purple) curve is the one-field solution, while the top (orange) curve is the two-field solution. For both plots we have set $\epsilon = 1/(16 \pi^2 )$. Note that the one-field solution is only valid for integer charge, while the two-field solution is valid for third-integer charge.}
\label{fig13}
\end{figure}

The two-field solution has no analytical expression. We therefore plot the numerical values for the chemical potential and scaling dimension for each solution in Figure~\ref{fig13}. However, taking the exchange symmetric limit, the second solution becomes tractable:
\be
\Delta^{(2)} \simeq \begin{cases} 
          3q & \epsilon\gap q \ll 1 \\[3pt]
          \frac{6\pi^2}{\lambda_2} \left(3 \frac{\lambda_2 q}{8\pi^2} \right)^{4/3} & \epsilon\gap q \gg 1
       \end{cases} \qquad (\eta \to 0)\,,
\ee
where now we may have $q \in \mathbb{Z}/3$.

For both solutions the transition to large charge behavior occurs when $q \sim 1/\epsilon$ (in the figure we are taking $\epsilon = 1/(16 \pi^2 )$). The linear behavior is expected at small charge, when the anomalous dimensions are negligible. Naively (but see below), the two solutions seem to correspond to the families of operators $\phi_1^n$ and $\phi_2^n$, respectively. In the large charge limit $\epsilon\gap q \gg 1$, the scaling changes to the universal $\Delta \sim q^{4/3}$.

The dominant (minimal action) saddle at integer charge is given by the one-field solution. At fixed charge $q = n$, the one-field solutions unambiguously correspond to the operators $\phi_1^n$. The scaling dimensions \eqref{sol1} are superadditive for integer charge, and therefore the conjecture \eqref{conv} holds in this model with $q_0=1$.

The same interpretation cannot be made for the two-field solution. The scaling dimension at small charge seems to draw a correspondence with the family of operators $\phi_2^{n}$, with $n = 3 q$. However, there are two problems with this conclusion. First, such an operator has the same classical dimension and charge as many other operators, for example the family $\partial^{2i} \phi_1^i \phi_2^{n-3i}$. All of these operators will in general mix together. Supposing that the two-field saddle is even associated to a particular operator, at large charge it is unclear what operator this would be.

The second problem is that this saddle simply has no corresponding `stable' CFT state in the following sense. By analytically continuing back to real time, one can see that both solutions are simply periodic orbits in configuration space. In fact, they are the unique spherically homogeneous orbits with fixed radial mode ($\rho_i$ constant). Furthermore, the charge-fixing boundary condition (\ref{eq: Saddle2}) turns out to be equivalent to the Bohr-Sommerfeld quantization condition for these orbits:\footnote{For the application of the Bohr-Sommerfeld condition to the neutral $\lambda \phi^4$ theory, see \cite{Antipin:2024ekk}.}
\begin{equation}\label{bs}
    \sum_a \oint \Pi_a \cdot \text{d}\xi_a \in 2\pi \mathbb{N} \, ,
\end{equation}
where the sum is over all fields $\xi_a$, with momentum conjugates $\Pi_a$, and the inner product is that of field space, \textit{i.e.}, $f \cdot g \equiv \int \text{d}^4 x \, f(x) g(x)$. Each stable Bohr-Sommerfeld orbit corresponds to a quantum state, demonstrating again that the small-action saddle corresponds to a state on the cylinder. However, the large-action saddle as noted above is situated at a local maximum with respect to $\omega$. Each independent perturbation around the orbit has an associated `stability angle.' An unstable direction implies an imaginary stability angle, reflecting the exponential growth of such modes. When \eqref{bs} is extended to one-loop, the result is a complex energy eigenvalue and thus a complex scaling dimension \cite{Rajaraman:1982is}.

An analogy can be made here with an unstable particle at infinite volume, where the propagator's complex pole is due to mixing with a continuum of states. Strictly speaking we are at finite volume and thus there is no decay. However, at large $q$ a near-continuum of states develops and the effect of giving an imaginary part to the energy is the same. This phenomenon is also reminiscent of gapped Goldstones, see \cite{Cuomo:2020fsb}.

Finally, note that every stable state on the cylinder will have an associated stable periodic configuration, but in general these will neither be spherically homogeneous nor have a constant radial mode, and will therefore be highly complicated (which is why we have not found them).

\subsection{Perturbations on the Semi-classical Configuration}\label{sec:pert}
We have found that the one-field saddles only describe the integer-charge states on the cylinder, denoted by $|n\rangle$ to reinforce their correspondence with the operators $\phi_1^n$. For general large non-integer charge $q \in \mathbb{Z}/3$, we are interested in finding the states $|q\rangle$ of minimal energy.

To obtain these states we can act on the states $| n  \ra$ with the operators $\phi_2$ and $\phi_2^\dag$. We will apply this logic to present a systematic treatment of perturbations. The Lagrangian \eqref{eq:CylinderPolarlagrangian} can be made time independent with the change of variables
\be
\label{eq:Phi2ToPhi}
\phi_2 \equiv e^{i \c/3} \varphi \, ,
\ee
where since $n_2=0$, we have $\chi = \chi_1$ (we also set $\rho \equiv \rho_1$). Note that since $\chi$ is $2\pi$ periodic, $\varphi$ is ill-defined. Another way to say this is that $\varphi$ (as well as $e^{i \c/3}$) is charged under a $\mathbb{Z}_3$ gauge symmetry, in the sense of the coset construction, see \eqref{eq:VarphiGTransform}. This gauge symmetry acts as
\be
\varphi \to h\gap \varphi\qquad \text{and}\qquad e^{i \c/3} \to h^{-1}\gap e^{i \c/3} \, ,
\ee
where $h \in \mathbb{Z}_3$ is a cube root of unity. This is the price we pay for time independence. The connection to the standard coset construction is made in Appendix~\ref{app:CCWZ}.

The Lagrangian becomes\footnote{We neglect the boundary term from \eqref{eff}, since it does not play a role in the following. However, note that because of this term the energies are actually associated with the effective Hamiltonian $H_\text{eff} = H - \mu Q$.}
\begin{align}\label{var}
\mc L = &\, \frac{1}{2} (\partial \rho )^2 + \frac{1}{2} \rho^2 (\partial \chi )^2 + \frac{1}{2} m^2 \rho^2 + \frac{\lambda_1}{16} \rho_1^4 \notag\\[3pt]
& + |\partial \varphi |^2 + \f{i}{3} \p_\m \c \l ( \varphi \p_\m \varphi^\dagger - \varphi^\dagger \p_\m \varphi \r ) + \f{1}{9}(\partial \chi )^2 |\varphi|^2 + \left( m^2 + \frac{\kappa}{2} \rho^2 \right) |\varphi|^2 \notag\\[3pt]
& + \frac{\lambda_2}{4} |\varphi |^4 + \f{\eta}{6\sqrt{2}} \rho \l ( \varphi^3 + \varphi^{\dagger 3}\r )\, .
\end{align}
Quadratic fluctuations about the small-action saddle corresponding to $\rho$ and $\c$ are identical to those discussed in~\cite{Badel_2019}. Therefore, we will focus on the quadratic fluctuations for $\varphi$ when $f_1 \neq 0$. The corresponding part of the Lagrangian is given by
\be
\mc L_\varphi^{(2)} = |\partial \varphi |^2 + \f{\m}{3} \l ( \varphi \dot \varphi^\dagger - \varphi^\dagger \dot \vp \r )  + \left( m_2^2 - \f{\m^2}{9} \right) |\varphi|^2 \, ,
\ee
where we used \eqref{eq: Saddle} and we have defined
\be
m_2^2 \equiv m^2 + \frac{\kappa}{2} f_1^2 \, ,
\ee
which will turn out to be the energy cost to excite a quantum of $\phi_2$ on the large charge saddle, as we derive in \eqref{eq:m2physics} below. Using \eqref{eq:muf} and taking the large charge limit, we have
\be
m_2^2 \underset{q \rightarrow \infty}{\simeq} \frac{2\kappa}{\lambda_1} \mu^2 \simeq 1.2 \, \mu^2 \, .
\ee
Introducing the canonical momenta\footnote{The factors of `$i$' here and below are due to the Euclidean signature. Similarly, $\pi$ and $\pi^\dagger$ are indeed Hermitian conjugate since $\dot{\varphi}^\dagger = - \dot{\varphi}$.}
\begin{subequations}
\begin{align}
\pi^\dagger & = i \f{\p {\mc L}^\varphi_{(2)}}{\p\dot \varphi} = i \Big( \dot \varphi^\dagger  - \f{\m}{3} \varphi^\dagger \Big ) \\[6pt]
\pi & = i \f{\p {\mc L}^\varphi_{(2)}}{\p\dot \varphi^\dagger} = i \Big ( \dot \varphi  + \f{\m}{3} \varphi \Big ) \, ,
\end{align}
\end{subequations}
we find the Hamiltonian density
\begin{align}\label{hamil}
\mc H_\varphi^{(2)} &\equiv \mc L^\varphi_{(2)} +i\pi^\dagger \dot{\varphi}+i\pi \dot{\varphi}^\dagger \notag\\[3pt]
&= \l | \pi - i \frac{\m}{3} \varphi \r |^2 + \bigg ( m_2^2 - \f{\m^2}{9} \bigg ) |\varphi|^2+|\nabla \varphi|^2 \, .
\end{align}
From this result, we immediately conclude that stability is realized if and only if $m_2\geq \mu/3$.

Expanding both $\varphi$ and $\varphi^\dagger$ in spherical harmonics and solving the equations of motion, we find the following spectrum of eigenmodes
\begin{subequations}
\be\label{omeg}
\omega = \pm \bigg( \f{\m}{3} \pm \sqrt{m^2+ \frac{\kappa}{2} f_1^2+J_\ell} \bigg)  \equiv \pm \bigg( \f{\m}{3} \pm \omega_\ell \bigg) \, ,
\ee
with
\be
J_\ell = \ell (\ell+d-2) \, .
\ee
\end{subequations}
The general mode expansion is given by
\begin{subequations}
\label{eq:phiModes}
\begin{align}
\varphi(\tau, \vec \theta\gap) & = \sum_{\ell,\vec m} \f{1}{\sqrt{2\omega_\ell}} \l ( a_{\ell \vec m} \, e^{-(\omega_\ell+\m/3) \tau} \, Y_{\ell \vec m}(\vec \theta\gap) + b^\dagger_{\ell \vec m}\, e^{(\omega_\ell-\m/3) \tau} \,Y^*_{\ell \vec m}(\vec \theta\gap) \r ) \\[3pt]
\varphi^\dagger(\tau, \vec \theta\gap)  & = \sum_{\ell,\vec m} \f{1}{\sqrt{2\omega_\ell}} \l ( b_{\ell \vec m} \,e^{-(\omega_\ell-\m/3) \tau} \,Y_{\ell \vec m}(\vec \theta\gap) + a^\dagger_{\ell \vec m} \, e^{(\omega_\ell+\m/3) \tau} \, Y^*_{\ell \vec m}(\vec \theta\gap) \r ) \, ,
\end{align}
\end{subequations}
where $\ell$ and $\vec m$ are angular momentum quantum numbers, and $a_{\ell \vec m}$ and $b_{\ell \vec m}$ are operators satisfying canonical commutation relations
\be
\big[a_{\ell \vec m},a_{\ell' \vec m'}^\dagger\big] = \big[b_{\ell \vec m},b_{\ell' \vec m'}^\dagger\big] = \d_{\ell\ell'}\d_{\vec m \vec m'} \, .
\label{eq:commutators}
\ee
The quadratic Hamiltonian \eqref{hamil} for the zero modes is positive definite as expected:
\be
H_\varphi^{(2)} = \Big ( m_2 + \f{\m}{3} \Big )a_0^\dagger\gap a_0 + \Big ( m_2 - \f{\m}{3} \Big )b_0^\dagger\gap b_0,
\ee
The non-homogeneous modes with $\ell > 0$ have an additional contribution from gradient energy.  So $a_0^\dag$ and $b_0^\dag$ create the lowest energy excitations with $q = \pm 1/3$.

Consistency with \eqref{eq:phiModes} implies that the creation and annihilation operators transform non-trivially under the $\mathbb Z_3$ gauge transformations. The spectrum of states with integer charge can therefore be constructed using $\mathbb Z_3$ singlet combinations of creation operators $a^\dagger_{\ell \vec m}$ and $b^\dagger_{\ell \vec m}$ acting on the ground state $| n \ra$. For instance, 
\be
a^\dagger_{\ell_1 \vec{m}_1} a^\dagger_{\ell_2 \vec{m}_2} a^\dagger_{\ell_3 \vec{m}_3} | n \ra\,, ~~ 
b^\dagger_{\ell_1 \vec{m}_1} b^\dagger_{\ell_2 \vec{m}_2} b^\dagger_{\ell_3 \vec{m}_3} | n \ra\,, ~~
a^\dagger_{\ell_1 \vec{m}_1} b^\dagger_{\ell_2 \vec{m}_2} | n \ra\,, ~~ \text{\it etc.}
\ee

To produce states with third-integer charge, we must act with an operator that carries charge $1/3$. An obvious guess is
\be
\mc O _{1/3} \stackrel{?}{=} e^{i\c /3}\,.
\ee
However, this operator is not a singlet under the gauge group $\mathbb Z_3$.  The correct gauge invariant operator is
\be
\mc O _{1/3} = e^{i\c/3} \varphi\,,
\ee
which is of course nothing but $\phi_2$. The corresponding energy levels can be extracted from the two-point function 
\begin{align}
\la n | \, \mc O _{1/3}^\dagger(\tau_2,\vec \theta_2) \, \mc O _{1/3}(\tau_1, \vec \theta_1) \, | n \ra &= e^{-(\mu/3) \, (\tau_2-\tau_1)}
\sum_{\ell, \vec m}\f{Y_{\ell \vec m}(\vec \theta_2) \, Y^*_{\ell \vec m}(\vec \theta_1)}{2\omega_\ell} e^{- (\omega_\ell - \mu/3) (\tau_2-\tau_1)} \notag \\[4pt]
&= 
\sum_{\ell, \vec m}\f{Y_{\ell \vec m}(\vec \theta_2) \, Y^*_{\ell \vec m}(\vec \theta_1)}{2\omega_\ell} e^{-\omega_\ell (\tau_2-\tau_1)}\,,
\end{align}
where we substituted $\c$ for its background \eqref{eq: Saddle} and also used \eqref{eq:commutators}, which is valid since we are working at leading order.
When the operators are well separated in Euclidean time, the zero mode dominates the sum and the physical meaning of $m_2$ becomes clear: it is the mass of $\phi_2$ on the saddle and the energy gap between the ground state $| n \rangle$ and the excited state $| n + 1/3 \rangle$. That is,
\be
\Delta(n + 1/3) = \Delta(n) + m_2 R \, , 
\label{eq:m2physics}
\ee 
where $\Delta(n)$ is given in \eqref{sol1}.

The states of charge $n+2/3$ are not so simple. We can create them by acting on $| n \rangle$ with $(\mc O _{1/3})^2$, in which case their energy is $2 m_2$. However, we can equally well act on $| n + 1 \rangle$ with $\mc O _{1/3}^\dagger$. Using \eqref{gap}, the energy difference between the states $\mc O _{1/3}^\dagger| n + 1 \rangle$ and $| n \rangle$ is $\mu + m_2$. Since $m_2 > \mu$, the latter state has lower energy and should therefore be identified as $| n + 2/3 \rangle$.  We conclude that
\be
\Delta(n + 2/3) = \Delta(n) + (\mu + m_2) R \, . 
\ee 
Given that the states $| n + 1/3 \rangle$ and $| n + 2/3 \rangle$ have energies above $| n + 1 \rangle$, the corresponding spectrum is not even monotonic, let alone superadditive. This was also the case at small charge, as we discussed at the beginning of Section~\ref{sec:TwoFieldModel}.  We conclude that we are forced to take $q_0 = 1$ in order for the superadditivity conjecture to hold in this model.

Note that these conclusions were driven by the fact that $m_2 > \mu$ at the fixed point.  Since the analog of this relation could be different in another model, we briefly comment on some general constraints on the parameter space. First, we note that the stability of the effective theory requires $m_2 > \mu/3$ for consistency. Another way to see this is by acting with $(\mc O _{1/3})^3$ on $| n \rangle$. If $m_2 < \mu/3$, this action would create a state with less energy that $| n + 1 \rangle$, contradicting the fact that this is the charge $n+1$ state of minimal energy.
This does not exclude the possibility that $\mu/3 < m_2 < \mu/2$. In this case, the state $| n + 2/3 \rangle$ would be created by $(\mc O _{1/3})^2$ and the spectrum would be fully monotonic. This is turn would imply a large charge restoration of microscopic superadditivity in the sense that (\ref{conv}) would apply for $q_0 = 1/3$ as long as at least one charge ($q_1$ or $q_2$) is large. The reason for this is that the scaling dimensions of order one charges are order one, and thus negligible when compared with those of large charges.

\section{Superadditivity From the Bottom Up}\label{eft}
Now we turn to a bottom-up analysis of superadditivity in the large charge  semi-classical EFT framework of Refs.~\cite{Hellerman_2015,Monin_2017}.  We will first analyze the minimal EFT, where the only light dynamical mode is the charge phonon, a Goldstone boson whose presence is mandated by the symmetry breaking properties of the semi-classical saddle.  We will review the well known result that this minimal EFT respects superadditivity. The rest of the section will be a step-by-step search for a superadditivity violating EFT, which  must obviously involve additional light degrees of freedom. Obtaining a self-consistent framework is not straightforward, but we will show that when the additional degree of freedom  is a genuine scalar dilaton, one can indeed find a consistent example that violates superadditivity. Of course our analysis can only test consistency within the EFT and cannot provide a concrete UV completion, and therefore our example may well live in the swampland.  The point of this section is simply to demonstrate that bottom-up EFT arguments alone cannot be used to prove the superadditivity conjecture. However, they can tell us what properties such a superadditivity violating theory must have.

\subsection{A Superadditivity Preserving EFT}
We here review  the simplest large charge EFT, outlining how satisfaction of superadditivity is just mandated by the very validity of the EFT framework.
Our focus here will be on the superfluid EFT, for the theory on the cylinder, that emerges on the large charge semi-classical saddle at leading order in the derivative expansion. As first pointed out in Ref.~\cite{Hellerman_2015}, and as further analyzed in \cite{Monin_2017}, the EFT has a non-linearly realized conformal invariance associated with the spontaneous symmetry breaking
\be\label{symbreak}
SO(d+1,1)\times U(1)_Q \to  SO(d) \times D'
\ee
where $D'=D-\mu Q$ is the effective unbroken dilatation generator (\emph{i.e.}, 
the effective Hamiltonian on the cylinder). This breaking pattern, when viewed on the cylinder, precisely corresponds to a conformal superfluid state, like that we already encountered in Section~\ref{micro}, and mandates the presence of a single Goldstone field. To write down the general conformally invariant theory, we can use the standard trick to first require both Weyl invariance and diffeomorphism invariance, and then take the metric to be non-dynamical. The Goldstone field $\chi$ transforms by a  shift under $U(1)_Q$ and as a Weyl singlet.\footnote{This second property is mandated  by the fact that $U(1)_Q$ commutes with the conformal group.} The  symmetry breaking pattern is then fully encapsulated in the background $\chi$ configuration 
\begin{align}
\chi = \mu\gap t + \pi\,.    
\label{eq:chimupi}
\end{align} 
Note that we will work with Minkowskian signature in this section. Weyl transformations act on the metric  $g_{\mu \nu}$  as
\be
g_{\mu \nu} \to e^{2 \sigma} g_{\mu \nu} \, ,
\ee
corresponding to Weyl weight $-2$. The combination  
\begin{align}
\hat{g}_{\mu \nu} \equiv (g^{\a \b} \partial_\a \chi \partial_\b \chi) g_{\mu \nu} = (\partial \chi)^2 g_{\mu \nu}\,,   
\label{eq:ghat}
\end{align} 
then constitutes a Weyl invariant metric, which we can employ to build the EFT action as an expansion in curvature:
\begin{equation}\label{gold}
    S = \int \text{d}^d x \, \sqrt{\hat{g}} \Big( c_1 + c_2 \hat{\mathcal{R}} + c_3 \hat{\mathcal{R}}_{\mu \nu} \hat{\p}^\mu \chi \hat{\p}^\nu \chi + \cdots \Big) \,,
\end{equation}
where $\hat{\mathcal{R}}_{\mu \nu}$ and $\hat{\mathcal{R}}$ are the Ricci tensor and scalar built out of $\hat{g}_{\mu \nu}$. Similarly, the index of $\hat{\p}^\mu$ has been raised with $\hat{g}_{\mu \nu}$. The dots represent terms at least quadratic in curvature. The dressed Ricci scalar is 
\be\label{r}
\hat{\mathcal{R}} = \frac{\mathcal{R}}{(\partial \chi)^2} - (d-1)(d-4) \frac{\nabla_\mu |\partial \chi|\, \nabla^\mu |\partial \chi|}{(\partial \chi)^4} - 2 (d-1) \nabla_\mu \l( \frac{\nabla^\mu |\partial \chi|}{|\partial \chi|^3} \r) \,.
\ee
Similarly for the Ricci tensor, we have $\hat {\cal R}_{\mu\nu} = {\cal R}_{\mu\nu} + {\cal O} (\p^2 \chi)$. On the cylinder $\mathbb{R} \times S^{d - 1}$, we have
\begin{equation*}
    \mathcal{R}_{i j} = \frac{d-2}{R^2} \delta_{i j}  \qquad\text{and}\qquad  \mathcal{R} = -\frac{(d-1)(d-2)}{R^2} \, , 
\end{equation*}
where $i,j$ are spatial indices and all other components of the Ricci tensor vanish.

From this point forward, we set the cylinder radius $R = 1$, so that the eigenvalues of the Hamiltonian on the cylinder
yield the spectrum of scaling dimensions.
Thus the action takes the form
\begin{align}
    S = \int \text{d}^d x \, \sqrt{g} \, |\partial \chi|^d \left[ c_1 - c_2  \frac{(d-1)(d-2)}{(\partial \chi)^2} + c_3 (d-2) \frac{(\partial_i \chi)(\partial_i \chi)}{(\partial \chi)^4} + {\cal O} (\p^2 \chi) \right] \, .\notag\\[3pt]
    \label{eq:SEFT}
\end{align}
Then the scaling dimension is simply given by the the Hamiltonian on the solution
\be\label{fourthirds}
H=\int \text{d}^d x \left (\dot \chi\frac{\partial {\cal L}}{\partial \dot \chi}-{\cal L}\right ) \,\,\,\Longrightarrow\,\,\, \Delta = \mu\gap q - L
\ee
given on the solution\footnote{This result matches that of Section~\ref{micro}, where the Legendre transform was associated with the  wave functions \eqref{bd}, fixing the charge at the past and future boundaries.}
\be\label{qdef}
\dot \chi = \mu\qquad\text{and} \qquad q = \frac{\partial L}{\partial \mu}  \, .
\ee
For this model, \eqref{qdef} implies
\be\label{qhere}
\frac{q}{\Omega_{d-1}} = c_1\gap d\gap \mu^{d-1} - c_2 (d-1)(d-2)^2 \mu^{d-3} \, .
\ee
Inverting the relation and plugging into \eqref{fourthirds}, we derive the relation between the operator dimension and the charge:
\begin{align}
\Delta &=  c_1 (d-1)\gap \Omega_{d-1}\gap {\rho}^\frac{d}{d-1} \l[ 1 + (d-2) \frac{c_2}{c_1}  {\rho}^{-\frac{2}{d-1}} + \mc O \l(\rho^{-\frac{4}{d-1}} \r) \r] \, , 
\label{spec}
\end{align}
with
\begin{align}
\rho = \frac{q}{c_1 d \, \Omega_{d-1}} \,.
\end{align}
Notice the Ricci tensor term of \eqref{gold}, proportional to $c_3$, does not contribute to the scaling dimension at tree level.
We see the leading $q$ dependence of $\Delta$ scales as the convex function $q^{d/(d-1)}$, and that higher curvature terms in \eqref{gold} lead to corrections suppressed by powers of $q^{-2/(d-1)}\sim \mu^{-2}$, as expected for higher derivative effects. These corrections, which threaten the superadditivity of \eqref{spec}, are small when the EFT is valid (we argue this better below). 

Concerning quantum  corrections (which we have and will continue to largely ignore in this work), the loop-counting parameter is
\be
\frac{1}{c_1 \mu^d} \sim \frac{1}{c_1\rho^{\frac{d}{d-1}}}\, ,
\ee
as  can be easily read off from the leading term in \eqref{eq:SEFT} (or seen more explicitly in \eqref{exps} below). This immediately tells us that the first quantum correction to \eqref{spec} is $\mc O (q^0)$ (for $d=3$ it is a constant, while for $d=4$ the vacuum energy is log-divergent so that the correction goes like $\log(q)$).

Now we analyze the region of validity for this EFT.  There are two ways the EFT can break down. The first is the standard breakdown of perturbative unitarity, which corresponds to the scale at which interactions become strong (quantum effects are large).  There is an additional source of breakdown due to the presence of higher-derivative terms.  Their inclusion causes the EFT to breakdown at the scale where these terms become of the order of the  kinetic term. To estimate both of these scales, we express \eqref{eq:SEFT} as a polynomial in the Goldstone field $\pi$:
\begin{align}\label{exps}
\mc L &\sim c_1 \mu^{d-2} (\p \pi)^2 + c_1 \mu^{d-3} (\p \pi)^3 + c_{2,3} \mu^{d-4} (\p^2 \pi)^2 \notag\\[3pt]
&\sim (\p \tilde{\pi})^2 + \frac{1}{(c_1 \mu^d)^{1/2}} (\p \tilde{\pi})^3 + \frac{c_{2,3}}{c_1 \mu^2} (\p^2 \tilde{\pi})^2\,,
\end{align}
where we have dropped $\mathcal{O}(1)$ factors and $c_{2,3} = \max(|c_2|, |c_3|)$. In the second line, we have canonically normalized the Goldstone kinetic term by substituting $\tilde{\pi} \sim (c_1 \mu^d)^{1/2} \pi$. In addition to the standard kinetic term, we have kept the leading interaction term and higher-derivative kinetic term, respectively.  We can therefore identify two scales that must be much larger than the curvature scale ($1/R=1$) in order for the EFT to be a sensible field theory:
\be\label{cutoffs}
\Lambda_\text{unitarity} \sim c_1^{1/d} \mu \gg 1  \qquad\text{and}\qquad \Lambda_\text{derivative} \sim \sqrt{\frac{c_1}{c_{2,3}}} \mu \gg 1 \, .
\ee
The cutoff of the EFT is then the minimum of the two. The first inequality obviously implies that quantum effects are small, so that they cannot compromise the superadditivity of \eqref{spec}. The second inequality implies that the first term of \eqref{qhere} dominates, so that $\mu \sim q^{1/(d-1)}$.  Another application of the second inequality to the second term in brackets in \eqref{spec} shows that this term is small. In other words, if the EFT of the Goldstone mode alone is valid, \eqref{spec} implies that the theory is superadditive.
 
Therefore, in order to accommodate violation of superadditivity  in the large charge regime we are forced to consider situations where the EFT entails some other light degree of freedom in addition to the hydrodynamic Goldstone $\chi$. In fact an indication that things may then work is offered 
by  supersymmetric examples (\emph{e.g.} Ref.~\cite{Hellerman_2017}) where the Goldstone comes with a scalar superpartner and where $\Delta \sim q$, which is only marginally superadditive.

\subsection{Superadditivity Violation From a New Degree of Freedom}

In this section, we show that adding a new light degree of freedom to the EFT can result in superadditivity violation.  This analysis will be restricted to the quadratic terms in the model, and so it will not be possible to fully discuss the EFT validity. 
In the next section, we will show how to write consistent interactions, which will justify the validity of the following analysis. We should also add that the same analysis limited to the quadratic action was carried out in Ref.~\cite{Orlando:2023ljh}, with an interpretation of the final result that slightly differs from ours.\footnote{That paper correctly concludes that $-\beta^2\mu^2\leq m_r^2< 0$ does violate convexity without any apparent instability, but then speculates, without any solid basis, that $m_r^2$ should in fact be positive.}

Let us consider the possibility that, in addition to  the fluctuation $\pi$ of the superfluid field ($\chi=\mu\gap t+\pi$), there exists another light mode $r$. The quadratic action  around the saddle point 
will in general have the form
\be\label{orl}
\mathcal{L} = \mathcal{L}_0 + \frac{1}{2} f^2 \dot{\pi}^2 + \beta f \mu \, r \dot{\pi} + \frac{1}{2} \dot{r}^2 - \frac{1}{2} m_r^2 r^2 \, ,
\ee
where $\mathcal{L}_0(\mu) = \mathcal{L}|_{\pi=r=0}$, $f$ is the scale of $U(1)$ symmetry breaking, the mass $m_r$ is assumed to be $<\mathcal{O}(1)$ and $\beta$ is a numerical coefficient expected to be $\mathcal{O}(1)$. We are only interested in the spherically homogeneous modes on the cylinder and have thus focused on the zero momentum modes. Due to the magnetic-like cross-term, the theory is stable as long as
\be\label{stab}
m_r^2 > -(\beta \mu)^2 \, .
\ee
This can be seen by considering the quadratic Hamiltonian:
\be\label{mass>0}
\mc H = \frac{1}{2 f^2} (\Pi_\pi - \beta f \mu r)^2 + \frac{1}{2} \Pi_r^2 + \frac{1}{2} m_r^2 r^2 \, .
\ee
Even though the Hamiltonian is unbounded from below when $m_r^2<0$, that does not necessarily signal an instability. That is because $\Pi_\pi$ is a conserved charge that  can be treated as an integration constant: the $r^2$ piece  arising from  the first term in \eqref{mass>0} should be then treated as an independent contribution to the effective mass of $r$, which explains \eqref{stab}.

To discuss convexity we need to study the relative increments of the charge $q\to q+\delta q$ and of the chemical potential $\mu \to \mu +\dot \pi\equiv \mu+\delta\mu$. 
Eq.~\eqref{orl}, together with the equations of motion for $r$ (on stationary configurations, \emph{i.e.}, satisfying $\dot r = 0$), implies
\be
\frac{\delta q}{\Omega_{d-1}}=\frac{\partial {\cal L}}{\partial \dot \pi}\bigg|_\text{on-shell}=f^2\dot \pi+\beta f\mu r= \left (1+\frac{\beta^2}{m_r^2}\right ) f^2\dot \pi\equiv \left (1+\frac{\beta^2}{m_r^2}\right ) f^2 \delta\mu \,.
\ee
We can therefore write
\be
\frac{\partial^2 \Delta}{\partial q^2} = \frac{\partial \mu}{\partial q}=\frac{\partial \delta \mu}{\partial \delta q}= \frac{1}{f^2 \Omega_{d-1}}\frac{m_r^2}{m_r^2+(\beta \mu)^2} \, .
\ee
If we take $m_r^2$ negative, but still subject to \eqref{stab}, the right hand side is negative, corresponding to a concave $\Delta$.  Therefore, this EFT violates superadditivity at large charge.  Of course, if we decouple $r$ by taking the limit  $m_r^2 \to \infty$, superadditivity is recovered 
in accordance with the general arguments presented in the previous section.

Since this Lagrangian does not include any interactions, we are not in a position to determine if this analysis is within the regime of validity of the EFT.  Next we will explain the rules for writing down an interacting EFT that violates superadditivity, and show that the cutoff is parametrically under control.

\subsection{A Marginally Superadditivity Violating EFT}

In this section, we seek an EFT of the symmetry breaking pattern \eqref{symbreak} with a matter field in addition to the Goldstone mode. In seeking a superadditivity violating theory, it makes sense to first look for one that is `marginally superadditive,' which is to say $\p^2\Delta / \p q^2 = 0$, or $\Delta \sim q$ at large charge. The relation \eqref{fourthirds} then implies that we should look for a theory where $\mu$ is a constant function of $q$.

A Lagrangian serving this purpose is 
\begin{align}\label{margEFT}
    \mathcal{L} &= \sqrt{\hat{g}} \left[ b_1 + \hat{\phi}^2 \left( 1 + \frac{N^2}{(d-1)(d-2)} \hat{\mathcal{R}} \right) + a_1 (\p \hat{\phi})^2 \right] \notag\\[6pt]
    &= |\partial \chi|^d \left[ b_1 + \hat{\phi}^2 \left( 1 - \frac{N^2}{(\partial \chi)^2} \right) + \cdots \right] \, ,
\end{align}
where the metric $\hat{g}$ is given in \eqref{eq:ghat}, $\hat{\phi}$ is a dynamical real scalar field with a hat to emphasize that it has Weyl weight $0$, and $N$ is a positive number. By rescaling $\hat{\phi}$, the coefficient of the second term has been fixed to unity. In the second line we have also truncated both higher derivative terms and the kinetic term for $\hat{\phi}$, since they will not affect the following analysis. 

On stationary and homogeneous configurations, $\hat{\phi}$ acts as a Lagrange multiplier.  Specifically, it fixes\footnote{We have used the suggestive symbol $N$ to indicate that $\mu$ can be made much larger than the curvature, despite the lack of another scale, due to some large $N$ in the UV theory. However, even in this limit the EFT will turn out to be flawed.}
\be\label{mu=N}
\mu = \dot \chi=N \, ,
\ee
This also means that $\hat{\phi}$ is massless at the saddle point. The equation of motion for $\chi$ on the other hand gives (also using \eqref{mu=N}):
\be\label{phiq}
\frac{q}{\Omega_{d-1}} = \left [b_1 d  + 2\hat{\phi}^2\right ] N^{d-1} \, .
\ee
By plugging the solution back into \eqref{fourthirds}, we find the scaling dimension:
\be
    \Delta = N \gap q - b_1\gap \Omega_{d-1}\gap N^d \, .
\ee
Since the leading order behavior is linear, its superadditivity depends entirely upon the sign of $b_1$:  only if $b_1 \geq 0$ is $\Delta$ superadditive.

Can we exclude $b_1 < 0$? Glancing at \eqref{margEFT}, one would superficially be lead to conclude that $b_1$ controls the kinetic term of Goldstone fluctuation $\pi$ and that it should  hence  be positive. But the $\pi$  kinetic term is in fact controlled by the second term in \eqref{margEFT}, given  $\hat\phi^2\propto q$ grows with $q$. This dominant contribution is strictly positive at large charge, implying that requiring a sound kinetic term sets no  preference for positive or negative $b_1$, 
and thus no preference for  super- or sub-additivity.

Having demonstrated that the Lagrangian in \eqref{margEFT} can violate superadditivity, it remains to understand if it defines a sensible EFT. It is straightforward to see that it \emph{does not} because the kinetic term for $\pi$ is dominated by the Ricci term, which involves unsuppressed higher derivative corrections. Explicit computation, focusing only on terms with time derivatives of $\pi$ for simplicity, indeed gives
\begin{equation}
    \mathcal{L} \supset N^2 \hat{\phi}^2 \left[ 5 \dot \pi^2 + \ddot \pi^2 \right] 
    \, .
\end{equation}
This shows that the derivative expansion breaks down at length scales of the order of the radius of the cylinder (recall $R=1$ here), in such a way that our construction does not make sense as a $d$-dimensional EFT. Note that this is the same problem that arises in the case of (\ref{gold}) if we take $c_1 \to 0$, and therefore shows that we cannot trust any conclusions about superadditivity derived from this construction.

This motivates finding EFTs that have the parametric flexibility demonstrated by this example but also  a  large separation of scales suppressing higher derivatives. Looking back at \eqref{margEFT}, we see that the higher derivatives came from $\hat{\mathcal{R}}$: the problem could be solved if we use a Weyl invariant metric that does not involve field derivatives instead of \eqref{eq:ghat}. That purpose can be achieved if the extra scalar is a genuine dilaton, which non-linearly realizes the conformal symmetry irrespective of the superfluid background. Consider then adding a  scalar $\phi$ with Weyl weight 1:
\be
\phi \to e^{-\sigma} \phi \, .
\ee
Now using $\phi$ and $(\partial \chi)^2$, we can build  an infinite family of Weyl invariant metrics.  For concreteness, let us consider, beside \eqref{eq:ghat}, 
\begin{align}
(\hat{g}_{0})_{\mu\nu} = \phi^2 \gap g_{\mu\nu}
\qquad \text{and}\qquad 
(\hat{g}_{1})_{\mu\nu} = |\partial \chi | \phi \gap g_{\mu\nu}\,. 
\end{align}
Taking from here on $d=4$ for simplicity, a  Lagrangian giving the same dynamics as (\ref{margEFT}) on homogeneous stationary solutions is then\footnote{Limited to the lowest derivative terms, the two Lagrangians are the same when making the identification $\phi = |\partial \chi | \hat{\phi}$.}  
\begin{align}\label{dileft}
    \mathcal{L} &= b_1 \sqrt{\hat{g}} + \sqrt{\hat{g}_1} + \frac{N^2}{(d-1)(d-2)} \sqrt{\hat{g}_0} \, \hat{\mathcal{R}}_0  \notag\\[5pt]
    &= b_1 (\p \chi)^4 + (\p \chi)^2 \phi^2 + N^2 \l[ (\p \phi)^2 - \phi^2 \r]    \, ,
\end{align}
where $\hat{\mathcal{R}}_0$ is the Ricci scalar built from $\hat{g}_0$. Again, note $R = 1$ here. Since $\hat{g}_0$ only depends on $\phi$, its curvature $\hat{\mathcal{R}}_0$ does not produce higher derivative terms for $\chi$. The cutoff of this EFT may only come from the scale of perturbative unitarity (since there are no higher derivative terms).  This cutoff is parametrically large as we will show below in Section \ref{secstrong} for a large class of models. We shall also demonstrate the stability of the theory.

This shows that one way to construct a valid EFT of the Goldstone boson with another light degree of freedom is if we identify this new mode with the dilaton. The tuning required to give the flat potential for such a mode can come for example from supersymmetry, see \emph{e.g.}\ \cite{Cuomo:2024fuy}.

\subsection{A Class of Strongly Subadditive EFTs}\label{secstrong}

The arguments of \cite{Cuomo:2024fuy} suggest that with the addition of a dilaton, an even broader set of behaviors should be possible.   By reinstating the explicit dependence on $R$, consider indeed the charge density $\rho$ and energy density $\varepsilon$ in a model where $\Delta \sim q^\beta$,  
\be
\rho = \frac{q}{R^{d-1}}  \qquad\text{and}\qquad \varepsilon = \frac{q^{\beta}}{R^{d}} \, .
\ee
Now, for $\beta < d / (d-1)$, the `macroscopic limit' $q\,,R\to \infty$ with $\rho$ fixed, results in a vanishing energy density. This in turn indicates the existence of a flat direction, and thus of  a dilaton. In the following we reverse the argument, using a dilaton to construct theories with $\beta < d / (d-1)$.

The addition of the dilaton to the EFT allows for an even broader range of behaviors, as we will now show.
Consider the following 4d theory with free parameter $\alpha$:
\begin{align}\label{lagstrong}
    \mathcal{L} &= \frac{1}{12} \sqrt{\hat{g}_0} \, \hat{\mathcal{R}}_0 + \sqrt{\hat{g}_{\a}} = \frac{1}{2} (\partial \phi)^2 - \frac{1}{2} \phi^2 + \phi^{4 - 2\alpha} |\partial \chi|^{2\alpha}\,,
\end{align}
where as before $\phi$ is a dilaton, and $\hat{g}_{\a} = |\partial \chi |^\a \phi^{2-\a} \, g$ with $0 \leq \a \leq 2$. Notice that for $\alpha = 1$ this is just (\ref{dileft}) with $b_1 = 0$ and $N = 1/\sqrt{2}$.
At the saddle point  this action implies
\begin{subequations}
\label{contsaddle}
\begin{align}
\mu &= \left[ \frac{1}{2} (2-\alpha)^{\alpha-2} \left( \frac{q}{\alpha \Omega_3} \right)^{\alpha - 1} \right]^{\frac{1}{1+\alpha}}
\label{contsaddlemu}
\\[3pt]
\phi_0 &= \left[ \frac{1}{2} (2-\alpha)^{2\alpha-1} \left( \frac{q}{\alpha \Omega_3} \right)^{2\alpha} \right]^{\frac{1}{2+2\alpha}} 
\\[3pt]
\Delta &\propto q^\frac{2\alpha}{1 + \alpha} \, .
\end{align}
\end{subequations}
We see that when $\alpha = 2$ and the Goldstone and dilaton decouple, one finds the expected $q^{4/3}$ behavior. Likewise, at $\a = 1$ we find the same `marginally convex' (linear) behavior as for the model, see \eqref{margEFT}.

However, for $\a < 1$, the dimension becomes a strictly concave function of the charge, thereby violating superadditivity. We will now discuss the region of validity for this EFT with $\a < 1$. 
 Expanding \eqref{lagstrong} around the saddle (\ref{contsaddle}), the zero-mode Lagrangian at quadratic order is
\begin{align}
\mathcal{L}^{(2)} &= \frac{1}{2} \dot{r}^2 - \frac{1}{2} [2(\alpha-1)] r^2 \notag\\[4pt]
&\hspace{12pt}+ \frac{1}{2} \big[2\a (2
\a-1) \phi_0^{4-2\a} \mu^{2\a-2}\big] \dot{\pi}^2 + \big[2\a (4-2\a) \phi_0^{3-2\a} \mu^{2\a-1}\big] \dot{\pi} r\, ,
\label{quadaa}
\end{align}
where we have set $\phi = \phi_0 + r$. The first immediate conclusion is that $\a > 1/2$ so that $\pi$ has a positive kinetic term. We also need to ensure that the negative mass squared $m_r^2 = 2(\a-1)$ is  compensated by the magnetic-like cross-term so that the theory is stable. For this model, the condition in \eqref{stab} becomes
\be
0 < m_r^2 + (\beta \mu)^2 = \frac{2(1 + \alpha)}{2\a - 1} \, .
\ee
For $\a > 1/2$, this effective mass is positive and \eqref{stab} is satisfied.

The solution in \eqref{contsaddlemu} implies that $\mu$ is a decreasing function of $q$ when $\a < 1$. Based on the intuition we developed from \eqref{cutoffs}, it is reasonable to worry that the cutoff of this theory also vanishes in the large charge limit. However, the presence of the dilaton introduces a new scale, and we must therefore revisit the analysis of the perturbative unitarity bound. Reading off the exponents directly from \eqref{contsaddle}, the theory around the saddle is an expansion in the dimensionless quantities
\be
\frac{\p \pi}{\mu} \sim \frac{\partial \pi}{q^\frac{\a-1}{1+\a}} \sim \frac{\partial \tilde{\pi}}{\left(q^\frac{\a}{2(1+\a)}\right)^2} 
\qquad \text{and} \qquad
\frac{r}{\phi_0} \sim \frac{r}{q^\frac{\a}{1+\a}}
 \,,
\ee
where $\tilde{\pi} \sim q^{1/(1+\a)} \pi$ is the canonically normalized Goldstone field, see \eqref{quadaa}. At the lower of these two scales, we expect UV degrees of freedom to become important (assuming a UV theory exists). 
 We therefore have
\be
\Lambda_\text{cutoff} \sim q^\frac{\a}{2(1+\a)}\,.
\ee
Since this cutoff grows as the charge $q$ increases, we conclude that the EFT expansion is well behaved in the regime of interest.

It is again interesting to consider the ``flat-space limit"  $R\to \infty$. In \eqref{lagstrong}, that corresponds to setting the second term, associate with curvature, to $0$. The equation of motion for $\phi$ then fixes $\mu = 0$. Except for $\a = 2$ (normal $\Delta \sim q^{4/3}$ scaling) and $\a = 1$ ($\Delta \sim q$ scaling of theories with a moduli space), the action  is singular at this point and the theory cannot be trusted. Therefore these new exotic scaling behaviors for $\Delta$ coincide with the non-existence of a flat-space limit.

The existence of a valid region of parameter space with $1/2 < \a < 1$ demonstrates that it is possible for a well to do  EFT to violate the superadditivity conjecture. However we must  emphasize that our model cannot be taken as a counterexample to the conjecture as it is not a complete CFT, but only  an EFT truncation.  It is well possible that  there exists no CFT  UV completing our EFT, meaning that  it lies in the `swampland'. While we do not seem to possess a criterion to assess that possibility, we can at least assess the plausibility of \eqref{lagstrong} on the basis of naturalness.  In fact, our result strongly relies on the absence of the dilaton quartic term $\lambda \phi^4 $ in the potential. In the absence of supersymmetry, this term is perfectly allowed by all the symmetries of the theory, a fact that makes the spontaneous breakdown of the conformal group to Poincar\'e  thoroughly non generic (see for instance \cite{Sundrum:2003yt,Coradeschi:2013gda}), and subject to the virtually infinite fine tuning $\lambda\to 0$. In a realistic non-supersymmetric theory,  it is possible to fancy very small $\lambda$ because of some tuning, but it seems hard to accept $\lambda$ exactly zero (except for in some large $N$ limit; see footnote \ref{footFlat}, as well as the discussion in \cite{Cuomo:2024fuy}). Adding  $-\lambda \phi^4$ to \eqref{lagstrong}, with $\lambda>0$ and small, and redoing our analysis one finds that $\Delta(q)$ is concave and convex in respectively the two ranges $q\lesssim 1/\lambda^{\frac{1}{3}+\frac{1}{2\alpha}}$ and $q\gtrsim 1/\lambda^{\frac{1}{3}+\frac{1}{2\alpha}}$. Indeed, as expected, in the second range the dilaton acquires a mass larger than $1/R$, whereupon it can be integrated out so that we fall back to the standard EFT dimension-charge relation $\Delta \sim q^{4/3}$. So it seems that for a generic Lagrangian, convexity is finally recovered for $q$ large enough. It is interesting that by putting together not only necessary criteria like stability or validity of the derivative expansion but also a plausibility criterion like naturalness, then superadditivity, at least asymptotically, seems inescapable. This comfortably matches the expectation from ordinary statistical mechanics, that a system cannot have an increasing chemical potential as the charge is decreased. 
 
On the other hand, the study of CFTs at large charge $q$ and large angular momentum $J$, shows that there exist regimes where $\Delta (q, J)$ is a concave function of $J$. To summarize, what we have learned here is that bottom-up arguments alone are not sufficient to prove the superadditivity conjecture.

\section{Outlook}\label{disc}

In this paper, we made a number of observations about the superadditivity conjecture, as applied in particular to the large charge regime of CFTs.  

We first clarified the workings of the semi-classical regime in a Wilson-Fisher model involving two fields of charge $1/3$ and $1$ respectively. We showed that the states of minimal dimension at given integer  charge are all controlled by the same family of trajectories, with only the integer charge field turned on. The minimal dimension states with third integer charge correspond instead to Fock excitations around those trajectories. 
Our results shows that, throughout the spectrum, from  the small to the large  charge  regime, superadditivity can be violated by taking $q_0=1/3$ in \eqref{conv}. As the integer charge states correspond to Fock excitations around semiclassical trajectories, we can in a sense say that superadditivity is only violated by quantum effects, which  involve small charges.

We then investigated the occurrence of subadditivity in the large charge regime for general CFTs, which, under broad circumstances, is expected to be described by a semiclassical saddle corresponding to a homogeneous superfluid state. The self-consistency of this result simply inheres in the self-consistency of the EFT for the superfluid dynamics around the saddle. We  reviewed the minimal situation where the EFT just involves the indispensable superfluid Goldstone boson, showing how superadditivity just follows from the very  validity of the EFT description, i.e. from  the validity of the derivative expansion. We then considered the situation where one additional light mode is present. By a trial and error approach, we were quickly driven to the realization that such a scenario can violate superadditivity in a self-consistent manner only if the new degree of freedom is a genuine dilaton field. We were thus able to construct a seemingly self-consistent EFT  that violates both superadditivity and convexity: the chemical potential is a decreasing function of the charge. Although we have not bothered to provide a formal proof  that a dilaton EFT  is the only option, we are rather confident that it is so. Physically the presence of such a genuine dilaton indicates that conformal symmetry is broken independently of the presence of a charge density. Without supersymmetry, this situation is known to be achievable only at the cost of  fine tuning, and thus appears to be rather implausible. Our result then shows that while no strict bottom-up technical proof of superadditivity in the EFT can be offered, the price one has to pay to build a counter-example seems too high to be met by any realistic (non-SUSY) CFT. Stated differently, the unnaturalness of our counter-example strongly suggests that it may well lie in the swampland, \textit{i.e.}, that there exists no complete CFT from which it descends as an EFT. That a genuine dilaton seems  necessary  to build a self-consistent EFT violating the superadditivity conjecture constitutes a sort of moral proof of the conjecture.

The case of supersymmetry requires a special analysis, which goes beyond the purpose of this paper. We must however point out  that, in a specific class of 3D supersymmetric models, Ref.~\cite{Cuomo:2024fuy} finds at the next-to-leading order in the inverse charge expansion 
$\Delta= a(q-\tilde q_0)$, with  $a$ and $\tilde q_0$ both positive, which  is  superadditive. This results fully relies on the negative sign of the term proportional to $\tilde q_0$, which corresponds to the Casimir energy. In that same reference, it is however argued that, beyond the specific class considered there,  there may plausibly  exist models where $\tilde q_0<0$,  which is instead subadditive. It would clearly be interesting to study this question and come up with a definite answer as to the sign  of $\tilde q_0$ in general. A counterexample with $\tilde q_0<0$ would still correspond to a violation of superadditivity associated to a finite quantum of charge, much like the violation associated with $q_0$ in \eqref{conv}.

There are many future directions that deserve to be explored.  It would be interesting to study different classes of CFTs involving fields with different charges in order to test the generality  of the above conclusions about the  semiclassical description of the spectrum. Are the violations of superadditivity always tied to Fock excitations around the trajectory? Can one have  situations where different saddles are dominant in different charge ranges, leading to first order transitions?
An obvious arena where to play this game is given by Wilson-Fisher fixed points involving multiple fields. Particularly interesting in this respect would be clockwork-like CFTs, involving a tower of fields with smaller and smaller charges. Unfortunately, as shown in Appendix \ref{app:proof}, it appears difficult to write down a Wilson-Fisher clockwork CFT, given that with the minimal set of scalars there exists no fixed point. In the spirit of \cite{Rychkov:2018vya,Osborn:2020cnf}, one could however attempt a more systematic study, with non-minimal field content. It would also be interesting to study the large semiclassical regime of the 3D Chern-Simons clockwork model of Ref.~\cite{Sharon_2023} to see, for instance, if a single family of trajectories dominates. As concerns instead the bottom-up EFT perspective, it may be worth to try to realize the  example of Section~\ref{eft} in a warped compactification. That would not offer a complete UV completion but would perhaps better illuminate the difficulties in providing one.

Understanding the boundaries of the swampland on the space of low energy theories continues to be an extremely active area of development.  The fact that we are unable to prove the superadditivity conjecture from low energy arguments alone should not surprise us.  It is after all a statement about the space of possible UV completions.  Nonetheless, our hope is that this paper clarifies and sharpens what this conjecture tells us about the space of EFTs that could never be UV completed.  Refining our understanding of the swampland will ultimately serve to inform our understanding of fundamental aspects of quantum field theory.

\acknowledgments

We are grateful to Jahmall Bersini, Gabriel Cuomo,  Eren Firat, and Stefan Stelzl for useful discussions. We would like to especially thank Gabriel Cuomo for providing multiple insights and for a careful reading of the manuscript.
TC is supported by the U.S.~Department of Energy Grant No.~DE-SC0011640. AG and RR are partially supported by the Swiss National Science Foundation under contract 200020-213104 and through the National Center of Competence in Research SwissMAP. The work of AM is supported in part by the National Science
Foundation under Award No.~2310243. RR acknowledges the hospitality of the Perimeter Institute for Theoretical Physics, the Center for Cosmology and Particle Physics at NYU and the Theory Division of CERN. RR also acknowledges support from the Simons Collaboration on Confinement and QCD Strings.

\appendix
\section*{Appendices}
\addcontentsline{toc}{section}{\protect\numberline{}Appendices}%

\section{Connection With the CCWZ Construction \label{app:CCWZ}} 

In general, the CCWZ construction~\cite{Coleman:1969sm, Callan:1969sn, Salam:1969rq} allows one to build an action once the symmetry breaking pattern is specified. Here, we will not describe the procedure in detail but will instead highlight the crucial steps relevant for the case at hand. For the symmetry breaking pattern\be
\mathbb R \times U(1) \to \mathbb R_\text{diag} \times \mathbb Z _3\,,
\ee
we introduce one Goldstone mode, denoted $\chi$ (details can be found in~\cite{Monin_2017}) 
\be
\Omega = e^{i \c}\,.
\ee
The field $\c$ parameterizing the coset space $U(1)/\mathbb Z_3$ does not span the full $U(1)$ group but is only defined modulo $2\pi / 3$. Its transformation under an element of $U(1)$ is not continuous
\be
g\gap \Omega = e^{i (\c + \a)} = e^{i \mathrm{mod} (\c+\a,2\pi/3)} e^{i \l[ (\c+\a)-\mathrm{mod} (\c+\a,2\pi/3)\r ]} 
\equiv e^{i\c'} e^{2\pi i n(\a,\c)/3}\,,
\ee
where the function $n(g,\Omega)$ is integer valued, hence non-differentiable. As a result, the covariant derivative
\be
\nabla_\m \c = -i \Omega^{-1} \p_\m \Omega
\ee
is not well defined. Instead, one has to use the covariant derivative for the field $3\c$, whose transformation under $U(1)$ is smooth.

Given a field $\vp$ transforming under a representation $\mathbb Z_3$, we can nonlinearly realize the action of the full $U(1)$ as
\be
\label{eq:VarphiGTransform}
g: ~ \vp' = e^{2\pi q n(g,\Omega) i/3} \vp\,.
\ee
The singularity (ambiguity) in the transformation \eqref{eq:VarphiGTransform} implies that this field is not well defined. Instead, only
\be
\phi = e^{i q \c} \vp
\ee
is well behaved and has the proper quantum numbers with respect to the broken group $U(1)$. Therefore, in building the Hilbert space, we have to select only singlets of the unbroken $\mathbb Z_3$. In particular, the covariant derivative 
$\nabla_\m \vp$ should be defined as
\be
\nabla_\m \vp = \p_\m \vp + i q \vp\nabla_\m \c \equiv e^{-i q \c} \p_\m \l ( e^{i q \c} \vp \r )\,.
\ee

\section{Non-existence of a Simple Clockwork Family \label{app:proof}}

In this appendix, we demonstrate that the na\"ive extension of the model \eqref{eq:lag} to a family of clockwork models fails. Consider the theory of $N$ complex scalar fields $\phi_i$, $i = 1 \dots N$, with charges $Q(\phi_i) = 3^{1-i}$.  The most general Euclidean Lagrangian is 
\begin{equation}
\label{eq:lagn}
    \mathcal{L} = \sum_{i}^N \bigg( |\partial \phi_i |^2 + \frac{\lambda_i}{4} |\phi_i |^4 \bigg) + \sum_{i< j=1}^N \kappa_{ij} |\phi_i |^2 |\phi_j |^2 + \sum_{i=1}^{N-1} \bigg(\frac{\eta_i}{6} \phi_i \phi_{i+1}^{\dag 3} + \text{h.c.}\bigg)\, ,
\end{equation}
where the Wilson coefficients are $\lambda_i$ with $i=1,\dots , N$, $\kappa_{ij}$ with $i,j=1,\dots , N$ and $i<j$ (for notational convenience we will take $\kappa_{ji} \equiv \kappa_{ij}$), and $\eta_i$ with $i=1,\dots , N-1$. Note that this reduces to \eqref{eq:lag} when we take $N=2$, $\eta_1 = \eta$, and $\kappa_{1 2} = \kappa$.
The one-loop fixed point equations for this theory are
\bae\label{neqs}
    \lambda_i &= 5 \lambda_i^2 + 4 \sum_{j\neq i} \kappa_{ij}^2 + 2 \eta_{i-1}^2 \\[3pt]
    \kappa_{ij} &= 4 \kappa_{ij}^2 + 2 \sum_{k\neq i,j} \kappa_{ik} \kappa_{jk} + 2 \kappa_{ij} (\lambda_i + \lambda_j) + \delta_{i+1,j} \eta_i^2 \\[3pt]
    \eta_i &= 3 (2 \kappa_{i,i+1} + \lambda_{i+1}) \eta_i \, .
\eae
These are really the rescaled couplings \eqref{barred} with the bar suppressed for simplicity (alternatively we have set $16 \pi^2 \epsilon = 1$). For the first equation to make sense in the case $i=1$, we have introduced $\eta_0 \equiv 0$.

Now we look for non-trivial solutions when $N > 2 $. The first equation in \eqref{neqs} implies that for all $i$,
\be\label{condd}
\lambda_i (1 - 5\lambda_i ) \geq 0 \, \implies \lambda_i \leq \frac{1}{5} \, .
\ee
Note that for stability $\lambda_i \geq 0$ as well. Next, using that non-trivial solutions must have $\eta_i\neq 0$, plug the final equation of \eqref{neqs} with $i=1,2$ into the equation for $\lambda_2$:
\begin{align}
0 &= -\lambda_2 + 5\lambda_2^2 + \l(\frac{1}{3} - \lambda_2 \r)^2 + \l(\frac{1}{3} - \lambda_3 \r)^2 + \sum_{j > 3} \kappa_{2 j}^2 + 2 \eta_1^2 \notag\\
&= -\frac{1}{216} + 6 \l(\lambda_2 - \frac{5}{36} \r)^2 + \l(\frac{1}{3} - \lambda_3 \r)^2 + \sum_{j > 3} \kappa_{2 j}^2 + 2 \eta_1^2 \notag\\  
&\geq \l( \frac{4}{225} - \frac{1}{216} \r) + 6 \l(\lambda_2 - \frac{5}{36} \r)^2 + \sum_{j > 3} \kappa_{2 j}^2 + 2 \eta_1^2 \, ,
\end{align}
where the last line follows from using \eqref{condd} for $i=3$. This line is however a sum of positive quantities, which violates the inequality. Therefore, there are no solutions when $N>2$.

While this is the simplest family of clockwork models, one could augment the family to allow for a multiplicity of fields of each charge. In such cases, the fixed point equations quickly grow quite complicated, and we have not checked if any solutions exist for this larger class of models.


\addcontentsline{toc}{section}{\protect\numberline{}References}%
\begin{spacing}{1.09}
\bibliographystyle{JHEP}
\bibliography{biblio.bib}
\end{spacing}

\end{document}